**TITLE PAGE** <u>**Original**</u>

(3) Title:

Reduction dependence of superconductivity in undoped T' cuprates

(4) Author's Names

Osamu Matsumoto[a], Aya Utsuki[a], Akio Tsukada[b], Hideki Yamamoto[c], Takaaki Manabe[d], *Michio Naito[a]

(5) Addresses

[a] Department of Applied Physics, Tokyo University of Agriculture and Technology

Naka-cho 2-24-16, Koganei, Tokyo 184-8588, Japan

[b] Geballe Laboratory for Advanced Materials, Stanford University,

Stanford, California 94305, USA

[c] NTT Basic Research Laboratories, NTT Corporation, 3-1 Morinosato-Wakamiya,

Atsugi, Kanagawa 243-0198, Japan

[d] National Institute of Advanced Industrial Science and Technology (AIST)

Higashi 1-1-1, Tsukuba, Ibaraki 305-8565, Japan

*) Corresponding author.   Address: Department of Applied Physics, Tokyo University of Agriculture and Technology, Naka-cho 2-24-16, Koganei, Tokyo 184-8588, Japan. Tel. +81 42 388 7229; fax: +81 42 385 6255.   E-mail address: minaito@cc.tuat.ac.jp.

# Reduction dependence of superconductivity in undoped T' cuprates

Osamu Matsumoto[a], Aya Utsuki[a], Akio Tsukada[b], Hideki Yamamoto[c], Takaaki Manabe[d], and *Michio Naito[a]

[a] Department of Applied Physics, Tokyo University of Agriculture and Technology

Naka-cho 2-24-16, Koganei, Tokyo 184-8588, Japan

[b] Geballe Laboratory for Advanced Materials, Stanford University,

Stanford, California 94305, USA

[c] NTT Basic Research Laboratories, NTT Corporation, 3-1 Morinosato-Wakamiya,

Atsugi, Kanagawa 243-0198, Japan

[d] National Institute of Advanced Industrial Science and Technology (AIST)

Higashi 1-1-1, Tsukuba, Ibaraki 305-8565, Japan

**Abstract**

We have recently achieved superconductivity in undoped T'-$RE_2CuO_4$ ($RE$: Pr, Nd, Sm, Eu, and Gd), using epitaxial thin films by metal organic decomposition. The key recipes to achieve superconductivity are low-$P_{O2}$ firing and subsequent vacuum reduction to minimize the amount of impurity oxygen atoms, which are very harmful to high-$T_c$ superconductivity. In this article, we report our investigation on the reduction dependence of superconductivity of T'-$RE_2CuO_4$. For thin films, the amount of remnant $O_{ap}$ atoms is difficult to evaluate but we propose that one good measure for this may be the $c$-axis lattice constant, which tells us whether the reduction is *insufficient* or *excessive*.

*PACS*: 74.10.+v, 74.25.Dw, 74.62.Bf, 74.72.Dn, 74.78.Bz

*Key words*: High-$T_c$ superconductivity, Mother compounds, Undoped superconductors, $Nd_2CuO_4$ structure, Apical oxygen

*) Corresponding author. Address: Department of Applied Physics, Tokyo University of Agriculture and Technology, Naka-cho 2-24-16, Koganei, Tokyo 184-8588, Japan. Tel. +81 42 388 7229; fax: +81 42 385 6255. E-mail address: minaito@cc.tuat.ac.jp.

**Introduction**

Very recently we have reported superconductivity in undoped T'-$RE_2CuO_4$ ($RE$: Pr, Nd, Sm, Eu, and Gd) [1,2]. The $T_c$ reaches ~ 33K in $Nd_2CuO_4$, which is the highest ever reported for T' cuprates. The specimens have been prepared in thin-film forms by metal organic decomposition (MOD). In order to achieve superconductivity in undoped T'-$RE_2CuO_4$, low-$P_{O2}$ firing and subsequent low-temperature vacuum reduction are prerequisite. This process minimizes the amount of impurity apical oxygen ($O_{ap}$) atoms left in the lattice, which are very harmful to high-$T_c$ superconductivity. However, it turned out that prolonged reduction degrades the film properties, and eventually makes films transparent and insulating with the *T'* structure preserved [1,3], implying that removal of $O_{ap}$ proceeds with loss of oxygen (O1) indispensable for superconductivity in the $CuO_2$ planes. Hence the process window (phase field) to obtain superconducting films is rather narrow, especially in the case of $Eu_2CuO_4$ and $Gd_2CuO_4$.

The most scientific way to optimize the $O_{ap}$ removal process with O1 intact is to rely on the data of the site-specific occupancy for $O_{ap}$, O1 and also O2 in the fluorite $RE_2O_2$ layers, as a function of temperature and oxygen partial pressure. Such data, in principle, are obtainable by neutron diffraction experiments [4,5], but comprehensive data are unavailable at present. Furthermore not only thermodynamics but also kinetics, namely oxygen diffusion, have to be taken into account in the low-temperature processes; the size and shape of specimens (*e.g.*, bulk vs film) will significantly affect the process parameters. In this article, we propose an alternative method to optimize superconductivity in undoped T'-$RE_2CuO_4$. Our method relies on the empirical trend that the *c*-axis lattice constant ($c_0$) increases with the amount of $O_{ap}$ atoms. For each *RE*, we plotted the $T_c$ and resistivity of all films prepared with different firing and reduction recipes as a function of $c_0$, then we found behavior universal to all *RE*, namely a crossover from a *insufficiently* to *excessively* reduced state. Based on this trend, we derived the optimal $c_0$ value for superconductivity.

**Experimental**

The superconducting $RE_2CuO_4$ thin films were prepared by MOD using *RE* and Cu naphthenate solutions. The details of our MOD method are described elsewhere [1]. Briefly, the stoichiometric mixture of naphthenate solutions was spin-coated on SrTiO$_3$ (STO) (100) or DyScO$_3$ (DSO) (110) substrate [6]. The coated films were first calcined at 400°C in air to obtain precursors, then fired at 850°C – 875°C in a tubular furnace under a mixture of O$_2$ and N$_2$, controlling the oxygen partial pressure $P_{O2}$ from 4 x 10$^{-5}$ atm to 2 x 10$^{-3}$ atm. Finally the films were reduced in vacuum (< 10$^{-4}$ Torr ≈ 10$^{-7}$ atm) at various temperatures ($T_{red}$ = 420 ~ 600°C) and time ($t_{red}$ = 5 ~ 60 min) for O$_{ap}$ removal. The film thickness was typically 800 Å. The lattice constant was determined from the peak positions in $\theta$-$2\theta$ scans in X-ray diffraction.

**Results & Discussion**

Figures 1 show the $\rho$(300 K) and $T_c$ as a function of $c_0$ for all Pr$_2$CuO$_4$ films prepared on DSO substrates with different $T_{red}$ and $t_{red}$. The $\rho$(300K) decreases rapidly as the $c_0$ decreases from the bulk value (broken line) [7], and has a minimum around $c_0$ ~ 12.18 Å. A further decrease of $c_0$ leads to a gradual increase of $\rho$(300K). We can divide a range of $c_0$ into the following three regions.

Region I: *insufficient* reduction (12.23 Å ≥ $c_0$ ≥ 12.20 Å)

The $\rho$(300K) decreases rapidly with decreasing $c_0$. Superconductivity with $T_c^{onset}$ ~ 25 K suddenly appears for $c_0$ < 12.22 Å. Decreasing $c_0$ towards 12.20 Å, the $T_c^{onset}$ gradually increases to ~ 30 K, and the $T_c^{end}$ also improves.

Region II: *optimum* reduction (12.20 Å ≥ $c_0$ ≥ 12.19 Å)

The $\rho$(300K) decreases gradually with decreasing $c_0$, and optimum superconductivity is obtained at $c_0$ ~ 12.195 Å with $T_c^{onset}$ > 30 K and $T_c^{end}$ > 27 K.

Region III: *excessive* reduction (12.19 Å ≥ $c_0$ ≥ 12.16 Å)

The $\rho$(300K) still decreases with decreasing $c_0$ until 12.18 Å, but the $T_c^{onset}$ starts to decrease and the superconducting transition becomes broad with $T_c^{end}$ below 4.2 K. With a further decrease of $c_0$ below 12.18 Å, the resistivity gradually increases and superconductivity eventually disappears.

Figure 2 shows the typical temperature dependences of resistivity, $\rho$(T), in the three regions. The $\rho$(T) of film A in region I shows an upturn at low temperatures, which is a feature specific to *insufficient* reduction. Film B in region II shows the best superconductivity. Further reduced film C in region II is even more metallic but with reduced $T_c$. Film D in region III shows only a trace of superconductivity but is still metallic in the whole temperature range with no low-temperature upturn, a feature that distinguishes *excessive* reduction from *insufficient* reduction.

The behavior observed in Figs. 1 and 2 can be explained by assuming that the two effects, namely $O_{ap}$ removal and O1 loss, are involved in the reduction process.

The predominant effect in region I is removal of $O_{ap}$, which improves the film properties. In contrast, the predominant in region III is loss of O1, which degrades the film properties, although $O_{ap}$ may be removed further in region III as judged from the further shrinkage of $c_0$ [8]. This interpretation means that the removal of $O_{ap}$ is slightly quicker than the loss of O1, which realizes the situation with $O_{ap}$ atoms mostly cleaned up but with O1 atoms mostly preserved, and provides a stage potential for high-$T_c$ superconductivity as in region II.

$Nd_2CuO_4$ and $Eu_2CuO_4$ show behavior similar to $Pr_2CuO_4$ as seen in Figs. 3. The $c_0$ value optimal to superconductivity for all *RE* is summarized in Table I. The region showing a full superconducting transition (filled circles) is narrower in $Eu_2CuO_4$ as compared with $Pr_2CuO_4$ or $Nd_2CuO_4$. The window is the narrowest in $Gd_2CuO_4$, in which the superconducting samples were obtained but with poor reproducibility even by nominally identical sample preparation. One can explain this *RE* dependence from a solid-state-chemistry point of view. As explained above, the two requirements, (1) to preserve O1 and (2) to clean up $O_{ap}$, have to be satisfied for superconductivity to appear. The *RE* dependent solid-state chemistry tells us that the O1 binding energy is higher and the $O_{ap}$ binding energy lower for a larger *RE* ion. The former trend can be derived from the *RE* dependence of the enthalpy of formation for T'-$RE_2CuO_4$ (-329.0 kJ/mol for Pr, -298.3 kJ/mol for Nd, -300.1 kJ/mol for Sm, -240.6 kJ/mol for Eu, and -234.4 kJ/mol for Gd) [9-12]. The latter trend is deducible from the thermogravimetric analysis (TG) by Zhu and Manthiram on the $O_{ap}$ desorption process for T'-$RE_2CuO_4$ [13-15], which demonstrated that the desorption temperature of $O_{ap}$ is lower for a larger *RE* ion. Both indicate that the requirements for superconductivity are more difficult to satisfy for a smaller *RE* ion.

We have also attempted to remove $O_{ap}$ for $Gd_2CuO_4$ films by a reduction process close to thermal equilibrium, namely by lowering $T_{red}$ and increasing $t_{red}$ to see any improvement in reproducibility, but the results turned out to be worse. Figures 4 show the XRD patterns of $Gd_2CuO_4$ films reduced in vacuum at 260°C – 300°C for 13 hours. For comparison, the XRD pattern of an optimally reduced film ($T_{red}$ = 440 °C, $t_{red}$ = 10 min, $T_c^{onset}$ ~ 19 K) is also included in Figs. 4. As $T_{red}$ is increased from 230°C to 300°C, the (00<u>10</u>) peak shows no smooth shift, but a jump from 81.0° (as-grown) to 81.9° at around $T_{red}$ ~ 270°C ($2\theta$ = 81° and 81.9°, corresponding to $c_0$ = 11.864 Å and 11.759 Å, respectively). A peak split is discernible at 256°C and 280°C, indicating the coexistence of *insufficiently* and *excessively* reduced portions with no *optimally* reduced portion in the films. The film with $T_{red}$ = 230°C was semiconducting whereas the film with $T_{red}$ = 300°C was transparent and insulating (extremely *excessive* reduction). In contrast, the superconducting $Gd_2CuO_4$ film shows a single peak, although somewhat broad, at $c_0$ = 11.805Å. This result indicates that it is difficult to clean up $O_{ap}$ atoms with O1 intact by the reduction process close to thermal equilibrium following thermodynamics, especially for smaller *RE*. The binding energies of interstitial $O_{ap}$ and regular O1 seem to be very close in T'-$RE_2CuO_4$.

**Summary**

We proposed that the *c*-axis lattice constant, $c_0$, is a quite useful parameter to optimize the reduction process to achieve superconductivity in undoped T'-$RE_2CuO_4$ since $c_0$ reflects the amount of remnant $O_{ap}$. The plots of $T_c$ and resistivity as a function of $c_0$ demonstrate a crossover from *insufficiently* to *excessively* reduced states, and tell us whether the reduction is *insufficient* ($O_{ap}$ still remains) or *excessive* (O1 goes out). Based on this trend, we obtained the $c_0$ value optimal for superconductivity. We also point out that the reduction process close to thermal equilibrium is not good for removing $O_{ap}$, but that the kinetics in reduction is quite important for optimizing superconductivity.

**Acknowledgements**

The authors thank Dr. Y. Krockenberger and Dr. J. Shimoyama for stimulating discussions, and Dr. T. Kumagai for support and encouragement. They also thank Crystec GmbH, Germany for developing new $RE$ScO$_3$ substrates. The work was supported by KAKENHI B (18340098) from Japan Society for the Promotion of Science (JSPS).

Table I. c-axis lattice constants of T'-$RE_2CuO_4$ for bulk and superconducting film samples. The bulk $c_0$ is taken from ref. 7.

| RE | $c_0$ (bulk) [Å] | $c_0$ (optimum film) [Å] |
|---|---|---|
| Pr | 12.234 | 12.197 |
| Nd | 12.163 | 12.121 |
| Sm | 11.972 | 11.938 |
| Eu | 11.903 | 11.855 |
| Gd | 11.881 | 11.805 |

Figure captions

Figures 1    Plots of $T_c$ (upper) and $\rho$(300 K) (lower) versus $c_0$ for all Pr$_2$CuO$_4$ films on DSO substrates prepared by MOD.  Upper: open and filled circles represent $T_c^{onset}$ and $T_c^{end}$.  Lower: filled circles, open circles and crosses represent films showing superconductivity with zero resistance, ones showing superconductivity without zero resistance, and ones showing no trace of superconductivity, respectively.  The dotted lines are guides for eye.  We divide a range of $c_0$ into the following three regions: *insufficient* (region I), *optimum* (region II), and *excessive* (region III) reduction.  The typical $\rho(T)$ data in the three regions (films A – D) are shown in Fig. 2.

Figure 2    Typical temperature dependences of resistivity of Pr$_2$CuO$_4$ films with *insufficient*, *optimum*, and *excessive* reduction.  Data A-D are from the corresponding films in Figs. 1.

Figures 3    Plots of $\rho$(300 K) versus $c_0$ for Nd$_2$CuO$_4$ films (upper) and Eu$_2$CuO$_4$ films (lower) prepared by MOD.  The meanings of symbols are the same as in Figs. 1.  The dotted lines are guides for eye.

Figures 4    XRD patterns of Gd$_2$CuO$_4$ films on STO reduced at $T_{red}$ = 260$^{\circ}$C – 300$^{\circ}$C for $t_{red}$ = 13h and optimally reduced film ($T_{red}$ = 440$^{\circ}$C, $t_{red}$ = 10 min, $T_c^{onset}$ ~ 19 K).  The patterns near the (00<u>10</u>) peak of T' are enlarged in the lower panel.

**Figure 1**

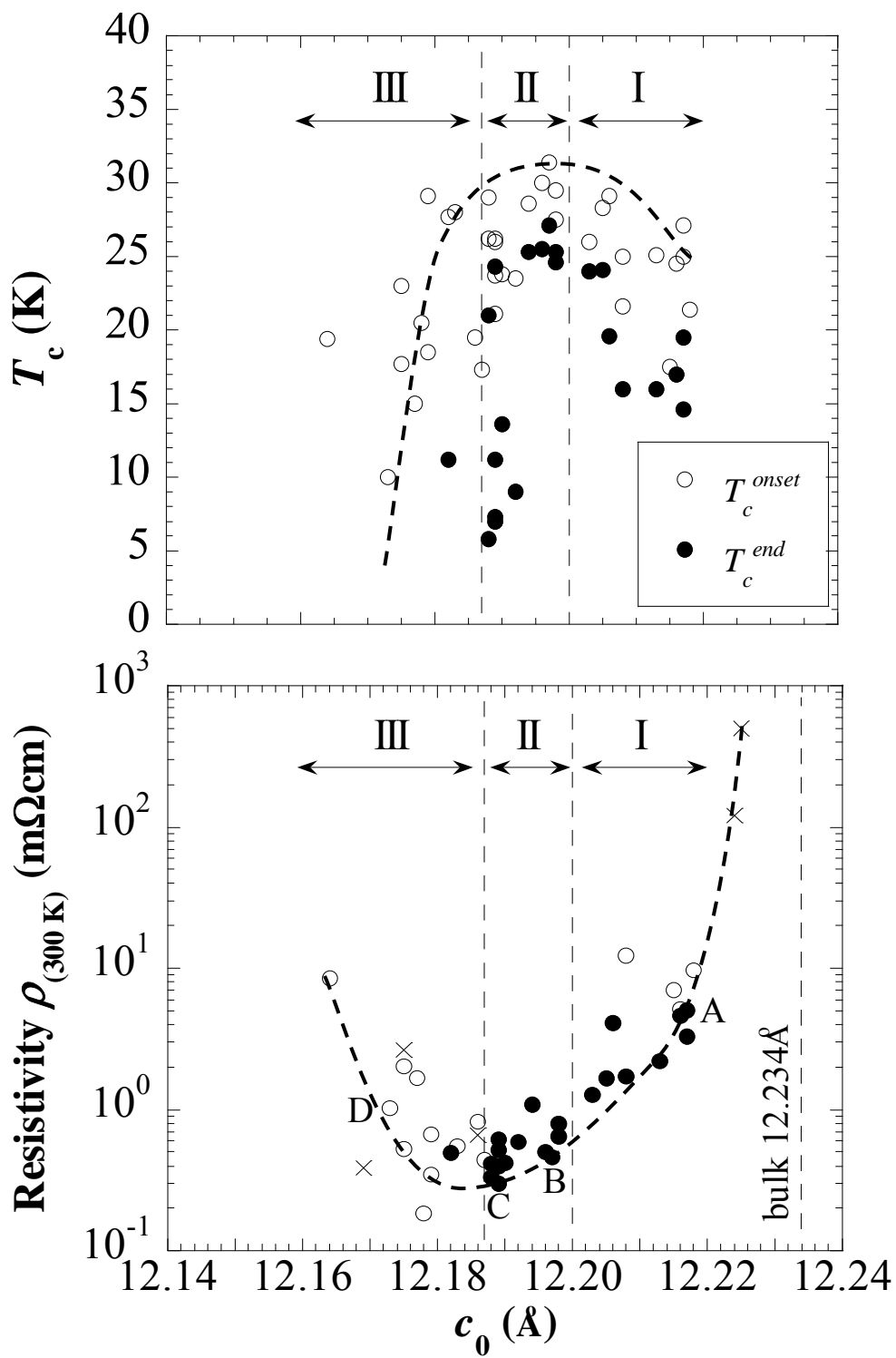

**Figure 2**

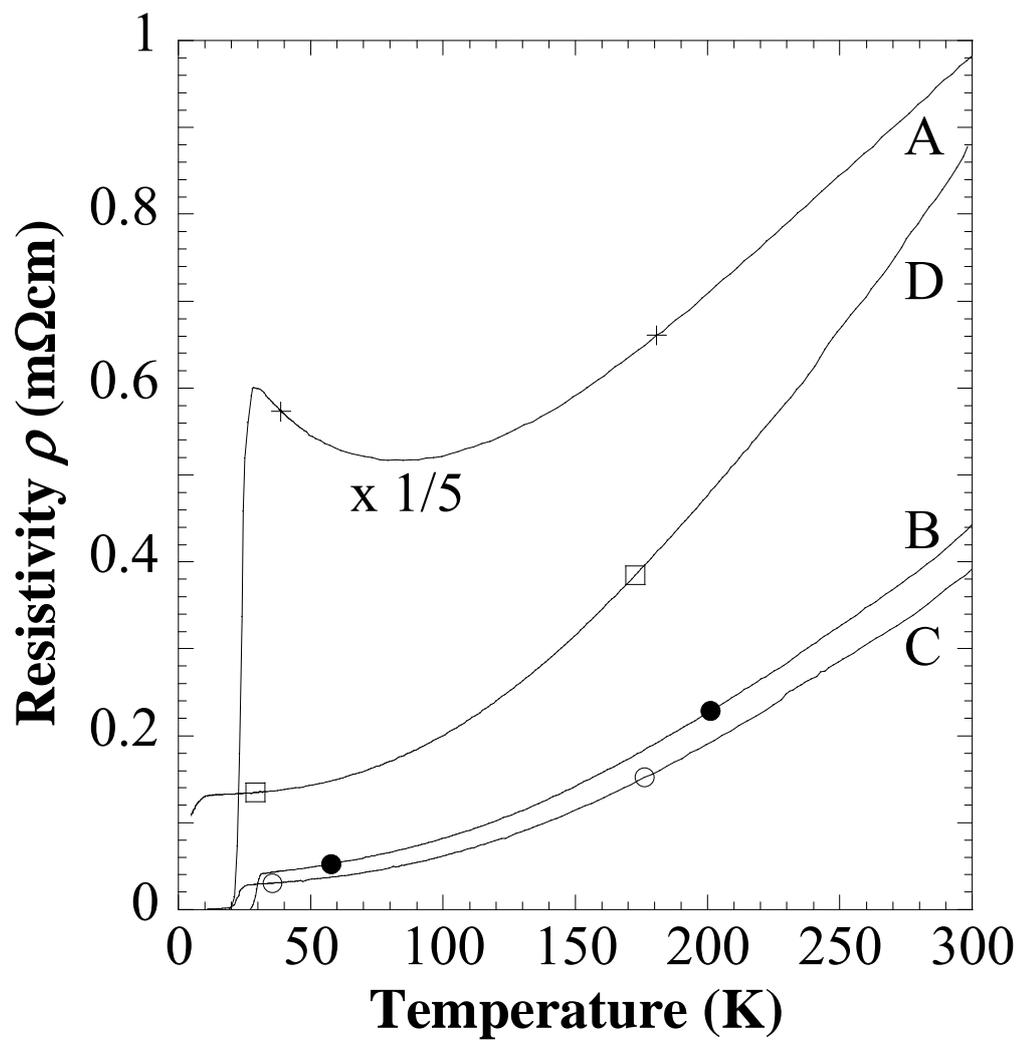

**Figure 3**

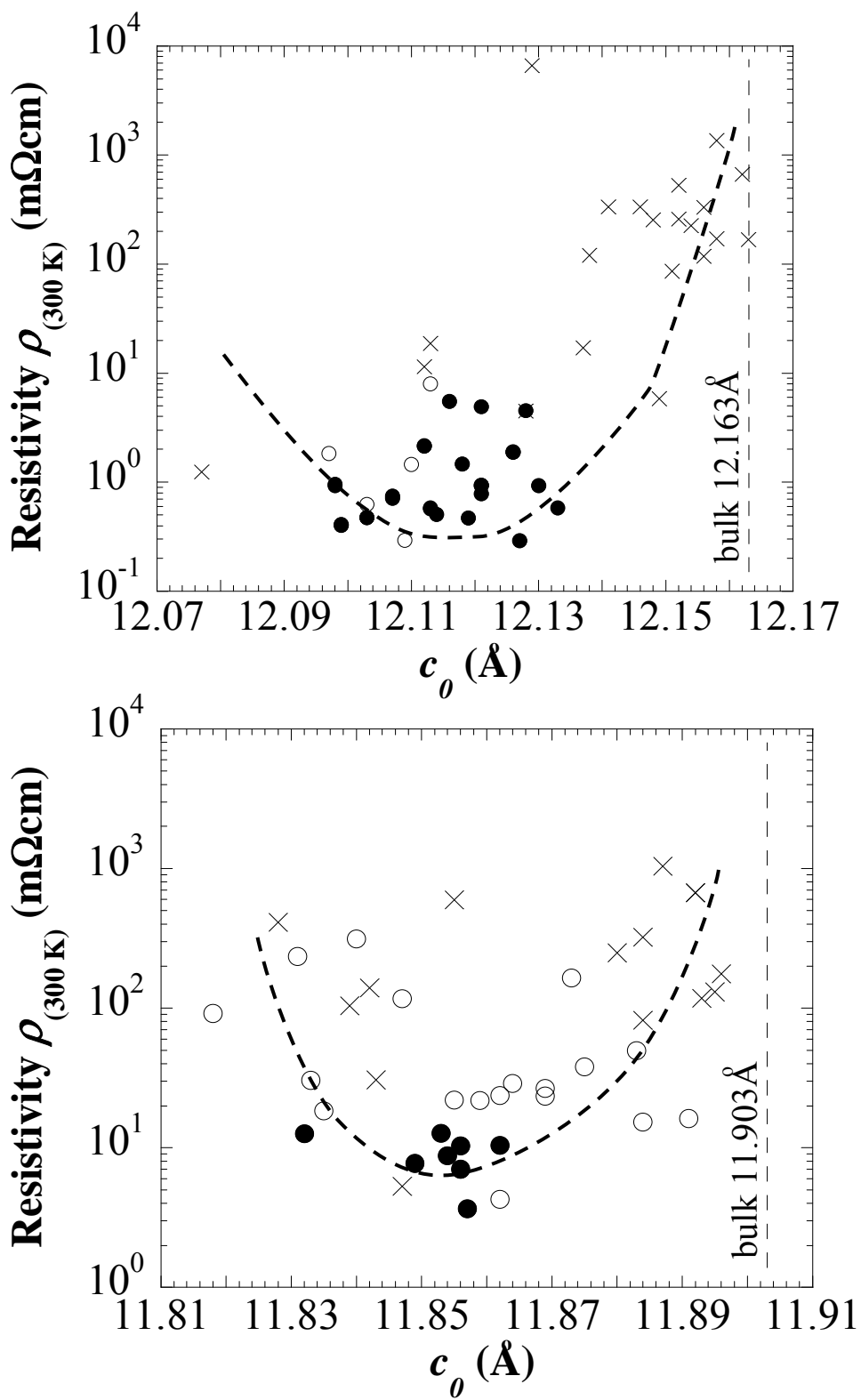

**Figure 4**

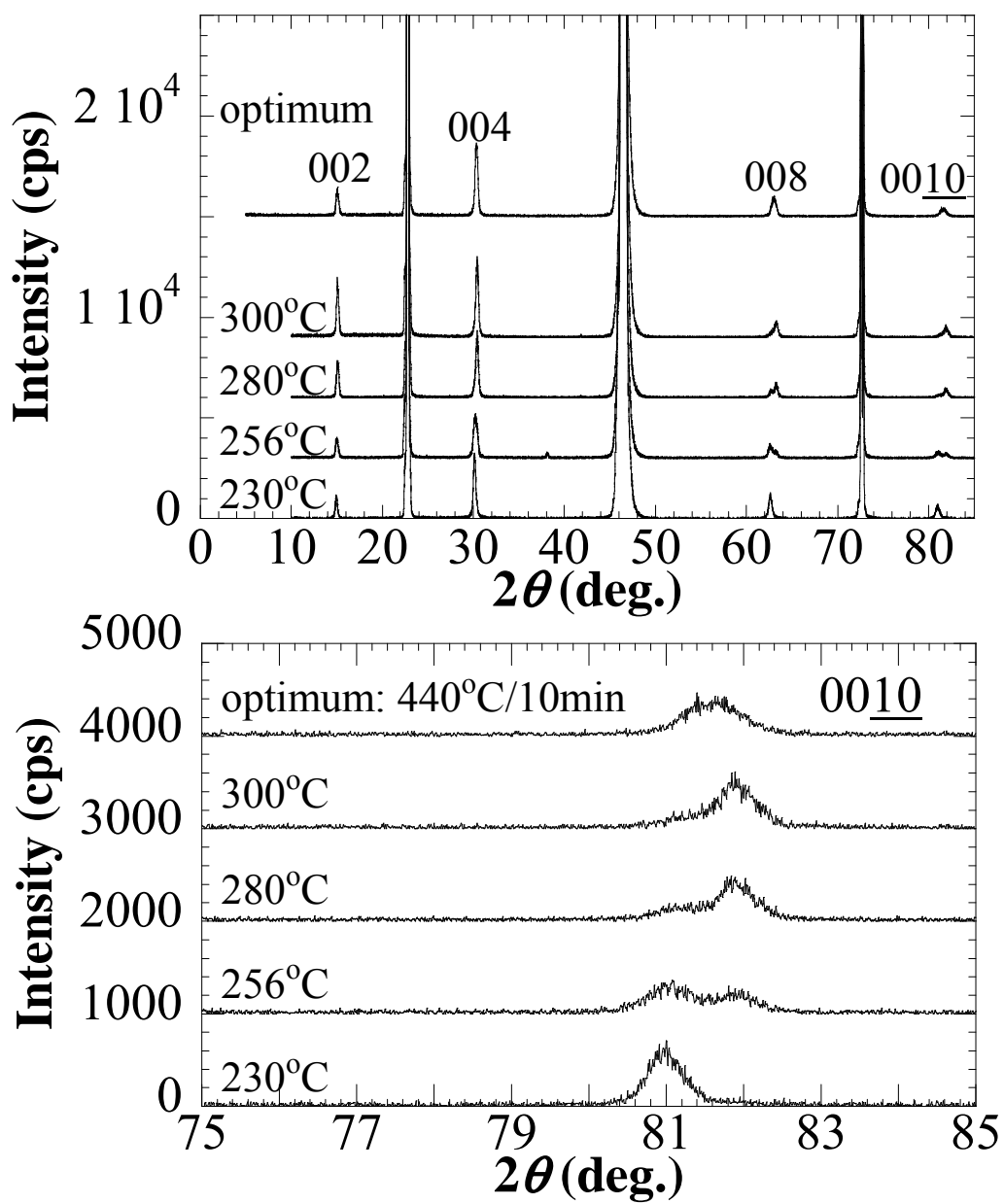